\documentclass[a4paper]{article}
\usepackage{graphicx}
\usepackage{fullpage}
\usepackage{physics}
\usepackage[numbers,sort&compress]{natbib}

\usepackage{amsmath}
\usepackage{braket}
\usepackage{color}
\usepackage{authblk}
\usepackage{comment}
\usepackage{bm}
\usepackage{soul}
\usepackage{xspace}
\usepackage{booktabs}
\usepackage{hyperref}

\def\trial {\Phi_{\rm trial}}
\def\evec {\Phi_{\rm FCIQMC}}
\def\HTC {\hat{H}_{\rm TC}}
\def\e {{\rm e}}

\newcommand{\avxz}[1]{aug-cc-pV{#1}Z}

\def\avdz {\avxz{D}\xspace}

\def\avtz {\avxz{T}\xspace}

\def\avqz {\avxz{Q}\xspace}

\newcommand{\neci}{\texttt{NECI}\xspace}
\newcommand{\molpro}{\texttt{Molpro}\xspace}
\newcommand{\casino}{\texttt{CASINO}\xspace}
\newcommand{\pyscf}{\texttt{pyscf}\xspace}

\newcommand{\pytchint}{\texttt{pytchint}\xspace}

\newcommand{\supercite}[1]{\textsuperscript{\citealp{#1}}}

\title{Transcorrelated Methods for Multireference Problems}
\author[1]{J. Philip Haupt\thanks{\href{mailto:p.haupt@fkf.mpg.de}{p.haupt@fkf.mpg.de}}}
\author[1]{Evelin M. C. Christlmaier}
\author[1]{Pablo L\'opez R\'ios}
\author[1]{Nikolay A. Bogdanov}
\author[1]{Daniel Kats}
\author[1,2]{Ali Alavi\thanks{\href{mailto:a.alavi@fkf.mpg.de}{a.alavi@fkf.mpg.de}}}
\affil[1]{Max Planck Institute for Solid State Research, Stuttgart, Germany}
\affil[2]{University of Cambridge, Cambridge, United Kingdom}

\date{}

\begin{document}

\maketitle
\begin{abstract}
    We apply the transcorrelated method to problems of multireference character. For this, we show that the choice of reference wavefunction during the Jastrow optimisation procedure is vital, and we propose a workflow wherein we use conventional multi-configurational methods to provide a reference wavefunction for Jastrow factor optimisation. This Jastrow function is subsequently used with transcorrelated-full configuration interaction quantum Monte Carlo within the xTC approximation (TC-FCIQMC) to yield highly accurate transcorrelated energies. This is demonstrated for N$_2$ using the aug-cc-pVTZ basis set, achieving chemical accuracy across the entire binding curve compared with experiment. We also apply the method to compute excitation energies of dinitrogen, CO and the ammonia molecule, where accurate results, comparable to the best available theoretical predictions, are obtained with modest basis sets.
\end{abstract}

\section{Introduction}

Accurate and efficient calculation of many-body correlation remains a central theme in theoretical chemistry. One particularly challenging aspect of this problem is the slow convergence of wavefunction ansatzes based on (anti-symmetrised) products of single-particle functions, due to the Kato cusp conditions.\supercite{kato_cusp_1957} Inclusion of the electron-electron distance $r_{12}$ into the wavefunction had already been observed in the 1920s to give quantitatively correct results.\supercite{slater_central_1928,hylleraasNeueBerechnungEnergie1929}
Among the explicitly correlated methods, the F12/R12 methods are particularly successful in accelerating convergence to the basis set limit, though these remain difficult to extend beyond the double-excitation level of coupled-cluster theory.\supercite{kutzelniggR12DependentTermsWave1985,kongExplicitlyCorrelatedR122012a}

Here, we pursue an alternative approach, dubbed the transcorrelated (TC) method, first proposed by Boys and Handy,\supercite{handy_be_1969,boysCalculationEnergiesWavefunctions1997} wherein a Jastrow correlator\supercite{Jastrow_1955} is used to similarity-transform the Born-Oppenheimer electronic Hamiltonian. The resulting operator 
is, however, non-Hermitian and contains three-body terms.

The TC method has recently enjoyed a resurgence of interest\supercite{umezawaTranscorrelatedMethodElectronic2003a,tsuneyukiTranscorrelatedMethodAnother2008b,ten-noFeasibleTranscorrelatedMethod2000a,sakumaElectronicStructureCalculations2006,leeStudiesTranscorrelatedMethod2023a,hinoApplicationTranscorrelatedHamiltonian2002,ammarExtensionSelectedConfiguration2022,ammarCompactificationDeterminantExpansions2024},  motivated partly by the availability of rather flexible forms of Jastrow factors which can be efficiently optimised 
and partly by the availability of powerful post-Hartree-Fock methods which can be used to solve the non-Hermitian eigenvalue problem.  These latter methods include full configuration interaction quantum Monte Carlo (FCIQMC), coupled cluster and density matrix renormalization group (DMRG).

Among these new studies is the finding that the FCIQMC\supercite{booth_fciqmc_2009,Cleland_initiator_2010} algorithm can handle the non-Hermitian character of the TC Hamiltonian.\supercite{Luo_tc_fciqmc_2018} Using VMC methods to optimise highly flexible Jastrow factors\supercite{haupt_optimizing_2023,lopezriosGeneric2012}, as well as recently-developed approximations\supercite{haupt_optimizing_2023,christlmaier_xtc_2023} in which the explicit three-body interactions of the TC Hamiltonian are eliminated, in this paper we extend these methods to solve strongly multireference problems such as the binding curve of the N$_2$ molecule to the fully stretched limit. We show that, in order to obtain high accuracy,  the Jastrow function needs to be optimised in the presence of a suitable multi-determinant reference function. We also apply the resulting methodology to compute  highly accurate excitation energies of N$_2$, NH$_3$ and CO with modest basis sets.

\section{Theory}

\subsection{Transcorrelation}

The theory behind the TC method has been extensively covered in previous publications\supercite{boys_handy_form_1969, Luo_tc_fciqmc_2018, cohenSimilarityTransformationElectronic2019a, christlmaier_xtc_2023}, and our purpose here is to cover aspects of the methodology which are particularly relevant to the present study.

The TC method relies on the so-called Jastrow ansatz,
\begin{equation}
\label{eq:sj}
    \Psi = \e^J\Phi,
\end{equation}
where $J$ is a real, symmetric, function depending on the spatial coordinates of all electrons and nuclei of the system, and $\Phi$ is an anti-symmetric wavefunction which we aim to solve for, using basis-set and full configuration interaction (FCI) type expansions, applied to the TC hamiltonian (which we review below).  In principle, for any $J$, there exists a corresponding $\Phi$ which would result in an exact solution $\Psi$ to the Schrodinger equation, but we expect a computational advantage (i.e $\Phi$ is simpler than $\Psi$) only if $J$ is chosen with the correct physics in mind. In this sense, we would like $J$ to reflect as much as possible the mathematical and physical characteristics of the Schrodinger problem, namely  universal cusp conditions (electron-electron and electron-nuclear) as well as system-specific dynamic correlation effects. With these  taken care of, we expect $\Phi$ to be computationally simpler to compute than $\Psi$.  Nevertheless, one is faced with the problem of how to obtain a suitable $J$ function in the first place. To date it has been done on the basis of minimising a variance of a \emph{Slater-Jastrow} wavefunction $e^J \Phi_\text{SD}$, i.e., the $\Phi$ function is replaced by a single-determinant wavefunction (usually the Hartree-Fock wavefunction) and the Jastrow function is optimised in the presence of this reference wavefunction for a fixed set of orbitals.\supercite{haupt_optimizing_2023}  This approach proves to be problematic in certain situations, which we aim to solve in this paper.

If Eq.\ \ref{eq:sj} is inserted into the Schrodinger eigenvalue problem and rearranged, we obtain the TC eigenvalue problem for $\Phi$:
\begin{equation}
    \label{eq:tc}
    \HTC \Phi = E \Phi,
\end{equation}
where $\HTC$ is the transcorrelated Hamiltonian, $\HTC = \e^{-J}\hat{H}\e^J$. This operator can be exactly re-expressed using the Baker-Campbell-Hausdorff formula:
\begin{align}
   &\HTC = \e^{-J}\hat{H}\e^J = \hat{H} + [\hat{H},J] + \frac{1}{2}[[\hat{H},J],J] \nonumber \\
    &= \hat{H}
    - \sum_{i}\left(\frac{1}{2}\nabla_{i}^{2}J
              + (\nabla_{i}J) \cdot \nabla_{i}
              + \frac{1}{2}(\nabla_{i}J)^{2}\right).
    \label{eq:bch}
\end{align}
Terms beyond the second commutator identically vanish for a Jastrow factor depending only on the spatial coordinates of the electrons, since only the kinetic energy operator $\hat{T}$ of the Hamiltonian gives rise to a non-zero commutator with $J$, but being a second-order differential operator, leads to a pure multiplicative function at the second-commutator level (the differential operators are exhausted); the subsequent commutators with $J$ must therefore vanish.  The last line of equation \eqref{eq:bch}  follows from the fact that $\hat{T}$ is a one-body operator $\hat{T}=-\frac{1}{2}\sum_i \nabla^2_i$.

If $J$ is further expressed as a sum over two-body terms:
\begin{equation}
       J=\sum_{i<j} u({\bf r}_i,{\bf r}_j),
\end{equation}
then $\HTC$ can be expressed as
\begin{equation}
\HTC = \hat{H} - \sum_{i<j} \hat{K}(\bm r_i, \bm r_j) - \sum_{i<j<k} L(\bm r_i, \bm r_j, \bm r_k), \label{eq:htc}
\end{equation}
where $\hat{K}$ is a non-Hermitian two-body operator, and $L$ is a Hermitian three-body operator. Expressions for these operators can be found, for example, in reference \cite{haupt_optimizing_2023}.

The most computationally expensive term of equation \eqref{eq:htc} is the three-body term $L$. However, effective approximation schemes have been developed, such as neglecting three-body contributions,\supercite{haupt_optimizing_2023} or using the xTC approximation\supercite{christlmaier_xtc_2023} where upon neglecting explicit three-body contributions, the remaining three-body terms are ``folded'' into the lower-order terms via normal ordering with respect to a reference wavefunction, as described below.

The Hamiltonian is normal ordered with respect to $\Phi_{SD}$:\supercite{kutzelniggNormalOrderExtended1997,kutzelniggCumulantExpansionReduced1999,christlmaier_xtc_2023}
\begin{equation}
 \begin{aligned}
    H_N &= \HTC - \langle\Phi_{SD}|\HTC|  \Phi_{SD} \rangle \\
        &= F_N + V_N + L_N,
\end{aligned}
\end{equation}
where the one-, two-, and three-body operators are (using Einstein summation)
\begin{subequations}
    \begin{align}
        F_N =
        \bigg[&
            h_P^Q
            +\big(U_{PR}^{QS}-U_{PR}^{SQ}\big)\gamma^R_S \nonumber\\
            &-\frac{1}{2}\big(L_{PRT}^{QSU}-L_{PRT}^{SQU}-L_{PRT}^{USQ}\big)\gamma^{RT}_{SU}
        \bigg]\tilde a^{P}_{Q},
    \end{align}
    \begin{align}
        V_N =\frac{1}{2} \bigg[
            U_{PR}^{QS}
            -\big(L_{PRT}^{QSU}
            -L_{PRT}^{QUS}-L_{PRT}^{USQ}\big)\gamma^T_U
        \bigg]\tilde a^{PR}_{QS},
    \end{align}
    \begin{eqnarray}
        \label{eq:LN}
        L_N =-\frac{1}{6} L_{PRT}^{QSU}\tilde a^{PRT}_{QSU},
    \end{eqnarray}
\end{subequations}
with $U=V-K$, $\tilde a^{P\dots}_{Q\dots}$ being the normal-ordered
excitation operators, and $\gamma^{P\dots}_{Q\dots}=\langle \Phi_{SD}| a^{P\dots}_{Q\dots}|\Phi_{SD}\rangle $ being the density matrices. The correction to the zero-body term $E_\mathrm{nuc}$ is then the expectation value of the three-body operator,
\begin{equation}
    \bra{\Phi_{SD}}L\ket{\Phi_{SD}} = -\frac 16L_{PRT}^{QSU}\gamma_{QSU}^{PRT}.
\end{equation}

The higher-order density matrices are also approximated as the antisymmetrised products of the one-body density matrix. That is, $\gamma_{SU\dots}^{RT\dots}\approx\mathcal{A}(\gamma_S^R\gamma_U^T\cdots)$, where $\mathcal A$ is the antisymmetriser operator. This approximation is exact for single-determinant references, and the only density matrix needed in the xTC approximation is the one-body reduced density matrix (1RDM). Ignoring the three-body terms $L_N$ leads to an improved scaling with respect to basis set size, while directly contracting intermediates to the lower-body corrections obviates the necessity to calculate $L$ and results in excellent energy agreement.\supercite{christlmaier_xtc_2023}

\subsubsection{Application: N$_2$ Binding Curve}

A popular ``stress test'' for quantum chemistry methods is the binding curve of N$_2$. A highly accurate experimental binding curve\supercite{leroyAccurate2006} exists to serve as the``ground truth" over a wide range of N-N distances, allowing for a useful benchmark for theoretical methods. At equilibrium, this system is essentially single reference in character, i.e., the electronic wavefunction is dominated by a Hartree-Fock wavefunction, but as the bond is stretched, the system becomes strongly multireference, where multiple determinants make significant contributions to the ground-state wavefunction.

To see how well TC fares against such problems, consider the methodology outlined in Ref.\,\cite{haupt_optimizing_2023}. The Jastrow factor described there has the Drummond-Towler-Needs (DTN) form:
\begin{equation}
    \label{eq:jastrow}
    J = \sum_{i<j}^Nv(r_{ij}) + \sum_i^N\sum_I^{N_A}\chi(r_{iI})
      + \sum_{i<j}^N\sum_I^{N_A}f(r_{ij}, r_{iI}, r_{jI}),
\end{equation}
with
\begin{equation}
    \label{eq:dtn-jastrow-ee}
    v(r_{ij})    = t(r_{ij},L_v)
                    \sum_{k} a_k r_{ij}^k ,
\end{equation}
\begin{equation}
    \label{eq:dtn-jastrow-en}
    \chi(r_{iI}) = t(r_{iI},L_\chi)
    \sum_{k} b_k r_{iI}^k ,
\end{equation}
\begin{equation}
    \label{eq:dtn-jastrow-een}
    f(r_{ij}, r_{i}, r_{j}) = \,t(r_{iI},L_f) t(r_{jI},L_f)
    \sum_{k,l,m} c_{klm}
    r_{ij}^k r_{iI}^l r_{jI}^m ,
\end{equation}
and $t(r,L) = (1-r/L)^3\Theta(r-L)$. The parameters of this Jastrow are optimised by minimising the variance of the reference function, taken by default to be the Hartree-Fock wavefunction:
\begin{equation}
    \label{eq:varref-hf}
    \sigma_\mathrm{ref}^2 = \sum_{I\neq \mathrm{HF}}|\bra{D_I}\HTC\ket{D_\mathrm{HF}}|^2,
\end{equation}

Using this workflow to optimise $J$, together with the xTC approximation, we can apply the TC-FCIQMC method to compute the binding curve of N$_2$ with the \avtz basis set. (Details of the FCIQMC calculations such as walker number and time-step are provided in the supplementary data). The result is shown in Fig.\ \ref{fig:binding-dip}, together with the experimental curve of LeRoy et al \cite{leroyAccurate2006}, in which the long-distance limit is set to twice the experimental energy of the N atom ($E_{at}=-54.58920$ Ha). We observe a generally reasonable agreement between the theoretically computed TC-FCIQMC curve and experiment. There is however a subtle unphysical ``dip'' in the stretched region of the theoretical curve at about 6 bohr.  A similar artefact was also found in an TC-DMRG study\supercite{liaoDMRG2023}. A further unpleasant feature is that the long-distance asymptote of the TC-FCIQMC curve is not size-consistent, dropping about 10 mHa below twice the atomic energy of the N atom computed at the same level of theory (i.e., TC-FCIQMC/aug-cc-pVTZ) . Since the post-Hartree-Fock (FCIQMC or DMRG) treatments are essentially at the FCI level, this implies that there is something likely wrong with the Jastrow factors themselves in the stretched limit.

\begin{figure}[htbp]
    \centering
    \includegraphics[width=0.45\textwidth]{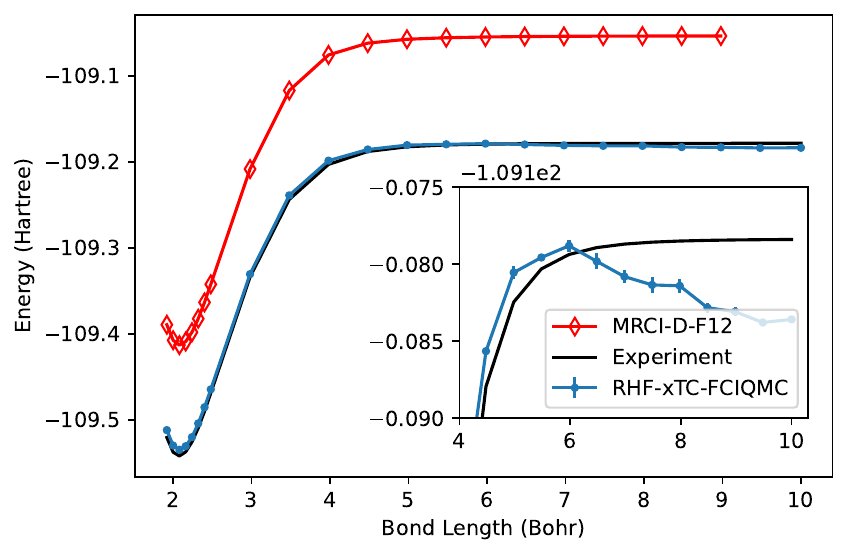}
    \caption{The TC-FCIQMC binding curve for N$_2$ with the \avtz basis set. An unphysical dip in the TC-FCIQMC calculation at large bond lengths is apparent when zooming in on the curve, as shown in the inset. For reference we also include the MRCI-D-F12 curve.}
    \label{fig:binding-dip}
\end{figure}

\subsection{Multireference TC Ansatzes}

Based on the discussion in the previous section, it is plausible that the transcorrelated workflow suffers from a single-reference bias, since the Jastrow factor is optimised in the presence of a single Slater determinant reference function:
\begin{equation}
    \label{eq:slater-jastrow-single-ref}
    \Psi_\mathrm{SJ} = \e^J\Phi_\mathrm{HF},
\end{equation}
which may be a poor representation of the wavefunction in multireference systems, thus negatively impacting the Jastrow optimisation and in turn the TC Hamiltonian $\HTC = \e^{-J}\hat H\e^J$.

We modify equation \eqref{eq:slater-jastrow-single-ref} to optimise for a multireference expansion,
\begin{equation}
    \label{eq:general-slater-jastrow}
    \Psi = \e^J\Phi_0,
\end{equation}
where $\ket{\Phi_0}=\sum_I c_I \ket{D_I}$. In practice, this modification results in two key changes in the workflow:
\begin{itemize}
    \item The objective function used during the VMC optimisation, equation \eqref{eq:varref-hf}, needs to reflect the multireference ansatz. In particular, it will need to be changed to
    \begin{equation}
        \label{eq:varref-md}
        \sigma_\mathrm{ref}^2 = \sum_{I} \big|\bra{D_I}\HTC\ket{\Phi_0}
         - \braket{D_I|\Phi_0}\bra{\Phi_0}\HTC\ket{\Phi_0}\big|^2,
    \end{equation}
    which is evaluated in VMC by sampling
    \begin{equation}
        S_\mathrm{ref}^2 =
          \frac 1 {n_\mathrm{opt}\!-\!1}
          \sum_{n=1}^{n_\mathrm{opt}}
            \left| \frac {\hat H({\bm R}_n) \Psi({\bm R}_n)}
                         {\Psi({\bm R}_n)} - {\bar E}_\mathrm{ref}
            \right|^2.
    \end{equation}
    \item The 1RDM, which is used in xTC during integration, must reflect this change as well. In particular, we use the 1RDM for the same state used in the Jastrow ansatz.
\end{itemize}

We propose two choices for $\Phi_0$, based on the premise that we wish the reference wavefunction to be qualitatively correct across the whole binding curve:
\begin{itemize}
    \item  A non-transcorrelated FCIQMC wavefunction. This approach quickly identifies the leading contributions to the wavefunction expansion at any geometry, and can correctly describe the strongly multireference stretched limit of the nitrogen molecule. Moreover, we can use the same HF orbitals (no further orbital optimisation), nor is an active space necessary to be identified. Since the purpose of this calculation is to identify a suitable reference wavefunction for subsequent Jastrow optimisation, the FCIQMC calculations can be terminated early  without reaching full convergence with respect to walker number\supercite{simulaEcp}. The 1RDM is calculated with replica method\supercite{overyUnbiased2014}.
    \item A complete-active-space-self-consistent-field CASSCF$(10e,8o)$ wavefunction, with a full valence active space.  By definition, in this approach the orbitals are optimised. An appealing feature of this wavefunction is that the CASSCF wavefunction describes correctly the purely static correlation in the stretched limit, making it naturally complementary to the role of the Jastrow factor. 

\end{itemize}

\subsection{Trial Wavefunctions in TC-FCIQMC}

Before we proceed to the main results of this paper, we mention here an important technical detail for accurately computing TC-FCIQMC energies in the presence of highly multi-reference wavefunctions, namely the use of appropriate trial wavefunctions\supercite{petruzielo_semistochastic_2012, bluntSemistochastic2015}.  
Conventionally, the projected energy is written
\begin{equation}
    E_{\rm proj} = \frac{\bra\trial\HTC\ket\evec}{\braket{\trial|\evec}}
\end{equation}
where $\ket\evec$ is the wavefunction according to the FCIQMC algorithm, and $\ket\trial$ is a fixed trial wavefunction with non-zero overlap with the exact solution. In practice, since the FCIQMC algorithm only provides stochastic snapshots of the wavefunction, a time-averaging is necessary. To this end, the numerator and denominator of the projected energy expression are time-averaged separately  (after a period of equilibration) and then the ratio is taken\supercite{bluntSemistochastic2015}. In order to provide a valid estimate of the stochastic error in this procedure, the serial correlation times of the time series of the numerator and denominator , as well as the co-variance of the two, must be determined. In this procedure, it is highly advantageous that the denominator is as large as possible,  and with small relative fluctuations, to prevent the denominator from approaching zero. This greatly reduces the effect of statistical fluctuations on the overall expression. In addition, note that if $\bra\trial$ is the left eigenvector of $\HTC$, then $E_{\rm proj}=E_0$ identically for any $\ket{\evec}$ and therefore it is also advantageous for the former to be as close as possible to the left eigenvector of $\HTC$.

In standard (non-transcorrelated) FCIQMC within the initiator approximation\supercite{Cleland_initiator_2010}, we take the leading $N_T$ determinants at some point of a calculation, form a subspace by constructing a $N_T\times N_T$ trial space $H_{\rm trial}$, diagonalise it exactly, and use the eigenvector as an approximation to the exact one, to be used as $\ket\trial$ in the calculation of the projected energy\supercite{bluntSemistochastic2015}. As $H$ is normally Hermitian, taking the left or the right eigenvector is irrelevant. For the TC hamiltonian, to get the left eigenvector in an equivalent way would involve doing a FCIQMC calculation on $\HTC^\dag$, but this is costly and instead we find taking the left eigenvector of the subspace from $\HTC$ to be a good approximation, as the two typically have similar leading determinants. Note also this method gives the correct energy as long as the trial wavefunction has nonzero overlap with the true eigenvector, so even if our choice is not perfect, it will still give the correct energy.

\begin{figure}[htbp]
    \centering
    \includegraphics[width=0.45\textwidth]{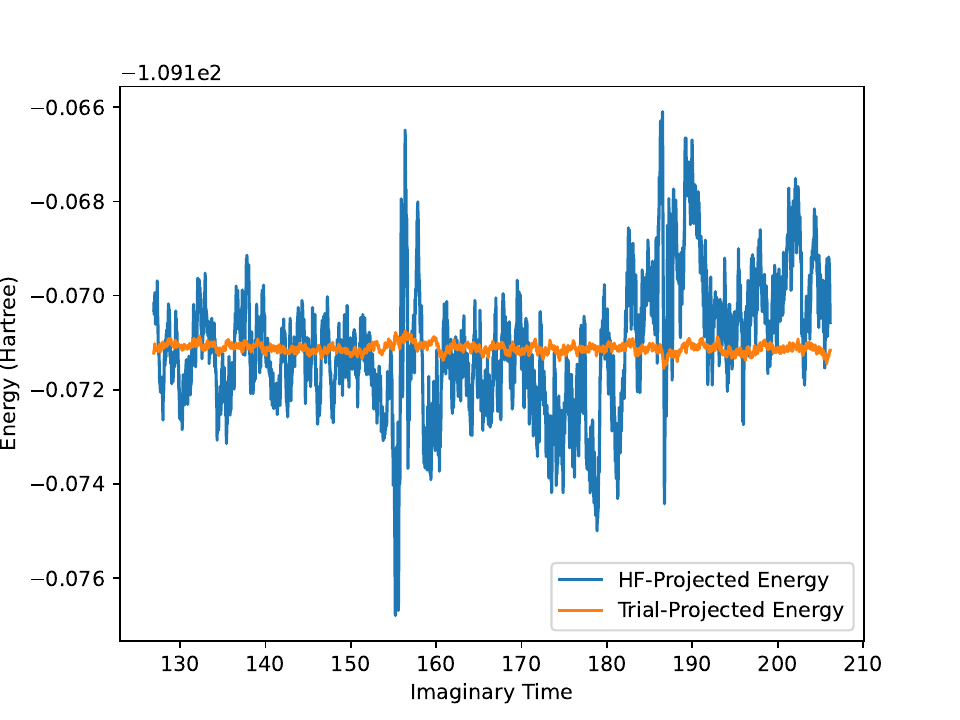}
    \caption{The HF-projected and trial-projected energy trajectories in imaginary time for N$_2$ with a separation of $10$ Bohr with the \avtz basis set and the Jastrow factor optimised for the variance of the FCIQMC energy. The calculation was done with $3\times 10^8$ FCIQMC walkers. This is a highly dissociated and hence multireference state. The trial-projected energy uses the top 20 determinants but substantially reduces the statistical fluctuations when compared to the HF projected energy.}
    \label{fig:trial_projected_energy}
\end{figure}

As an example of the effectiveness of this approach, consider a highly dissociated point in the binding curve of the nitrogen molecule. This is shown in figure \ref{fig:trial_projected_energy}, specifically at $10$ Bohr with the \avtz basis set and $10^8$ walkers. For this calculation, a Jastrow factor optimised with an FCIQMC-derived reference function was used.
We can see from this figure that using the trial-projected energy even with just a few determinants allows us to much more easily handle highly multireference problems. Thus, trial wavefunctions are used throughout the rest of this paper.

\section{Results and Discussion}

\subsection{Computational Details}

We calculate the energy of the nitrogen molecule across multiple bond lengths, ranging from $1.92$ to $10$ Bohr. We use the \avtz basis set,\supercite{avtz} which contains diffuse functions essential to describe the intermediate stretched region of the binding curve, while still being of relatively modest size. We calculate the non-TC CI vectors and 1RDMs for each geometry with FCIQMC (using \neci\supercite{Guther_neci_2020}) with the initiator approximation,\supercite{Cleland_initiator_2010} whereas those produced with CASSCF were done with \pyscf\supercite{pyscf1,pyscf2}. The CI vector is then used in the objective function for optimising the Jastrow factor with VMC using \casino.\supercite{Needs_casino_2020} Using the optimised Jastrow factor and 1RDM, we then calculate the relevant integrals for the xTC Hamiltonian using \pytchint.\supercite{tchint} Finally, TC-FCIQMC is performed on these integrals with \neci. Each geometry is calculated independently; that is, the Jastrow factor is optimised for each geometry separately. In order to keep memory usage manageable, we cut off the number of determinants in our CI vector to be $100$. Even at the dissociated limit, the number of relatively highly-weighted determinants is around $20$, so $100$ should be enough to capture the static correlation for the VMC optimisation.

Next, we also calculate the vertical excited-state energies using this workflow for dinitrogen, carbon monoxide and ammonia using Jastrow factors optimised with a CASSCF ansatz for $\Phi$. Excited states are also challenging multireference problems. We optimise the Jastrow factors in a state-specific manner. That is, for some excited state $\Phi_\mathrm{exc}$, our ansatz becomes $\Psi_\mathrm{exc} = \e^J\Phi_\mathrm{exc}$, thereby modifying our workflow slightly to optimise specifically for that state, as well as using its 1RDM for the xTC approximation. Naturally, the TC-FCIQMC calculation will be targeting this state. For cases where a triplet excited state of the same symmetry is lower in energy than a singlet excited state, the FCIQMC calculation will collapse to the triplet state. To overcome this, we use a spin-penalty term to target the singlet excited state.\supercite{weserSpinPurificationFullCI2022a}

For the N$_2$ binding curve, we compare against the highly accurate experimental curve determined in reference \cite{leroyAccurate2006}, and as a benchmark we compare against F12 calculations, performed in \molpro.\supercite{molpro1,molpro2,molpro3} For excited states we compare against extrapolated FCI calculations reported in reference \cite{loosMountaineering2018}.

Additional details for the calculations are included in the supplementary materials. The total energies of the TC-FCIQMC calculations presented in this paper can also be found there, as well as representative simulation trajectories. All FCIQMC calculations (transcorrelated and non-transcorrelated) have been performed with the initiator approximation using $n_\text{add}=3$.

\subsection{Revisiting the N$_2$ Binding Curve}

\begin{figure}[htbp]
    \centering
    \includegraphics[width=0.45\textwidth]{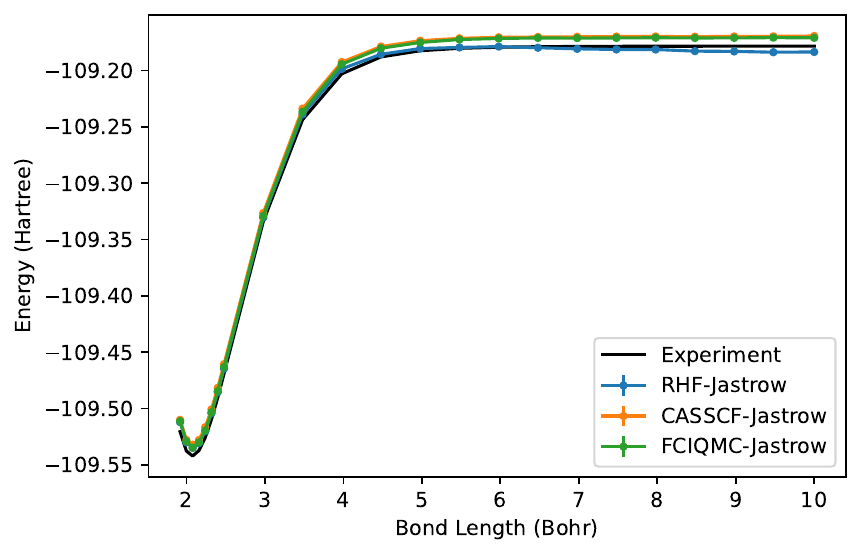}
    \includegraphics[width=0.45\textwidth]{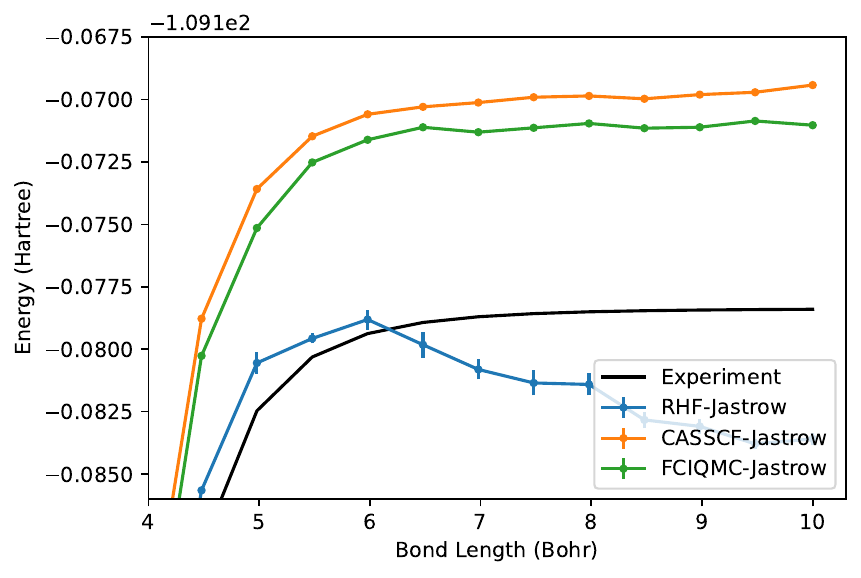}
    \caption{TC-FCIQMC energies for the nitrogen dimer for various points along its binding curve, between 1.92 and 10 Bohr radii. Calculations were performed with the \avtz basis set.
    }
    \label{fig:binding-curves-full-diss}
\end{figure}

As illustrated in figure \ref{fig:binding-curves-full-diss}, we obtain a qualitatively-correct binding curve for the nitrogen molecule using either of the multireference Jastrow factors. The remaining unphysical behaviour amounts to noise, which has a few potential sources:
\begin{itemize}
    \item The optimisation of the Jastrow factor is done with VMC, a stochastic algorithm. Since the optimisation is done independently for each point along the binding curve, this may lead to some noise.
    \item The FCIQMC calculations are also stochastic, and this may also lead to some noise, particularly in the dissociated limit when not using a multi-determinant trial wavefunction.
    \item In the case of the FCIQMC-Jastrow, even the non-TC calculation prior to Jastrow optimisation is stochastic. In this case, the CI vector and 1RDM collected from the non-TC-FCIQMC calculation is done so at only a snapshot in imaginary time, at the end of a FCIQMC calculation with $3\times 10^7$ walkers. One way to reduce this noise (not explored here) would be to average these values over imaginary time.
\end{itemize}
Thus, we can conclude that the biggest problem with the TC workflow has been resolved by introducing a multireference Jastrow optimisation (and 1RDM for the xTC approximation).

\begin{figure}[htbp]
    \centering
    \includegraphics[width=0.45\textwidth]{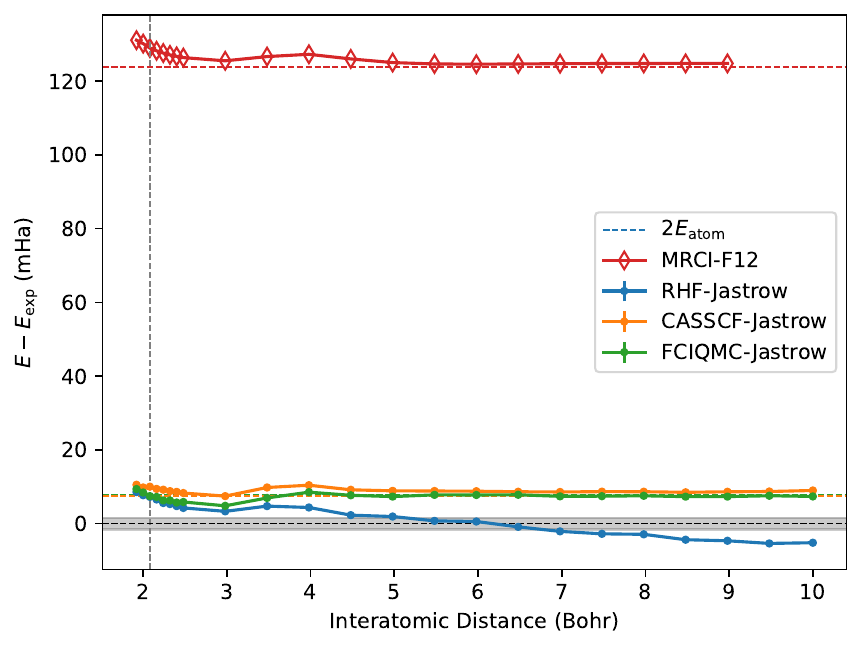}

    \caption{
        The difference between the theoretical binding curve and the experimental binding curve\supercite{leroyAccurate2006} for various methods.  Twice the atomic energy of the relevant method is presented as a dashed line in the same colour as the corresponding method. The dashed black horizontal line is at zero (i.e., the experimental value) with a region of $\pm 1.6$ mHa shaded grey. The grey vertical line indicates the equilibrium geometry of $2.08$ Bohr.
        From this plot, we see that the TC methods substantially improve upon the total energies,  whilst the two new Jastrow factors yield a much smaller non-parallelity error compared to the RHF-Jastrow curve.
    }
    \label{fig:binding-curves-experiment}
\end{figure}

To test the accuracy of our binding curves, we compare against the experimental binding curve,\supercite{leroyAccurate2006}. The difference $E-E_\text{exp}$ is plotted in \ref{fig:binding-curves-experiment}.
MRCI-D-F12 data is also included, which has been calculated with \molpro.\supercite{molpro1,molpro2,molpro3}
We find that TC substantially improves the total energies while the multireference Jastrow ansatz reduces the non-parallelity error by a substantial amount.

The dissociation energy, computed as the difference between twice the atomic energies and the energy of the equilibrium point in the binding curve, is presented in table \ref{tbl:size-inconsistency}, along with size consistency and non-parallelity errors. What we find is that both multireference TC approaches give accurate dissociation energies of experiment and both have small size consistency and non-parallelity errors, with the FCIQMC-Jastrow being particularly high-performing. The CASSCF-Jastrow is slightly less accurate, but has the advantage of being easily obtainable by several standard quantum chemistry packages, and is free from the stochastic considerations of the FCIQMC route.  For its ease of use, we will opt for it in subsequent work.

\begin{table}[htbp]
    \centering
    \begin{tabular}{lccc}
        Method & $E_\mathrm{diss}$ & $\epsilon_\mathrm{sc}$ & $\epsilon_\mathrm{NPE}$\\
        \hline
        TC-FCIQMC:\\
        ~~RHF-Jastrow & 364.0 & 12.7 & 8.0\\
        ~~CASSCF-Jastrow & 361.2 & -1.5 & 3.1\\
        ~~FCIQMC-Jastrow & 364.2 & 0.5 & 4.6\\
        \hline
        MRCI-D-F12 & 358.7 & -0.9 & 6.5\\
        HEAT\supercite{fellerSurvey2008}
        & 363.9 & -- & --\\
        Experiment\supercite{leroyAccurate2006}
        & 363.7 & -- & --
    \end{tabular}
    \caption{
    The TC-FCIQMC/aug-cc-pVTZ dissociation energy , size-consistency error $\epsilon_\text{sc}$ and non-parallelity error $\epsilon_\text{NPE}$ for the three different Jastrow factors applied to the dinitrogen binding curve. These quantities are calculated according to: $E_\mathrm{diss} = \sum_\text{at} E_\text{at} - E_\text{mol}(r=2.08 a_0)$ , $\epsilon_\mathrm{sc}=\sum_\mathrm{at}E_\mathrm{at} - E_\mathrm{mol}(r=10 a_0)$ and the $\epsilon_\mathrm{NPE}$ = Max error - Min error (relative to the experimental curve) respectively. All energies are in mHa. 
    We compare against MRCI-D-F12, as well as the high-accuracy HEAT benchmark from reference \cite{fellerSurvey2008} and experiment.
    }
    \label{tbl:size-inconsistency}
\end{table}

\subsection{Excited States}

One major advantage of this updated workflow is the ability to explicitly target specific states. We demonstrate this by calculating excited states for N$_2$, CO and NH$_3$ using state-averaged CASSCF Jastrow ansatzes with $(10e,8o)$, $(10e,8o)$ and $(8e,7o)$ active spaces, respectively. For each state, the Jastrow factor is optimised independently with the CI vector for that state. That is, $\Phi_0$ in equation \ref{eq:general-slater-jastrow} is set to the state-averaged CASSCF wavefunction for that state.

\begin{figure}[htbp]
    \centering
    \includegraphics[width=0.5\textwidth]{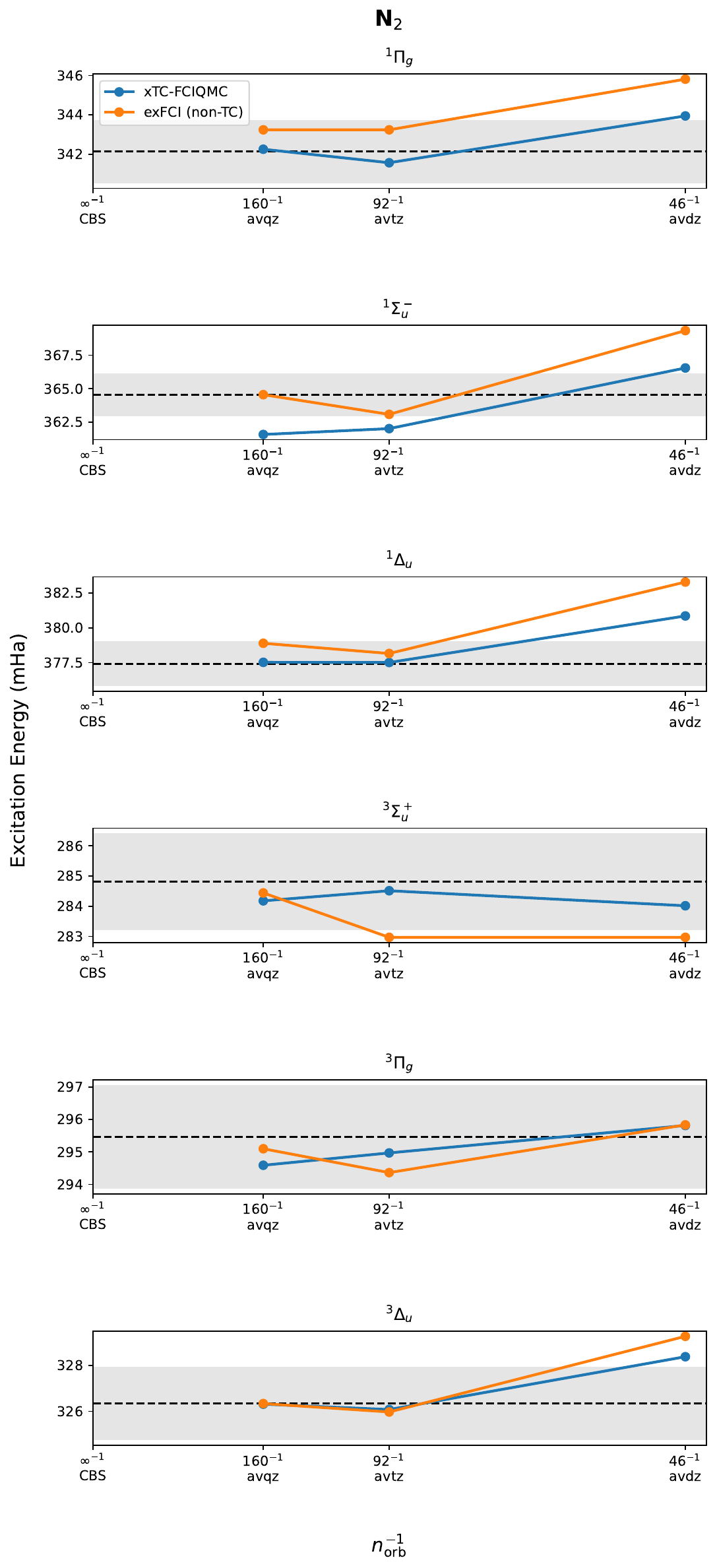}
    \caption{
        Excitation energies for dinitrogen compared with experiment\supercite{oddershedeComparison1985,huberConstants1979}.
    }
    \label{fig:excited-states-n2}
\end{figure}

\begin{figure}[htbp]
    \centering
    \includegraphics[width=0.5\textwidth]{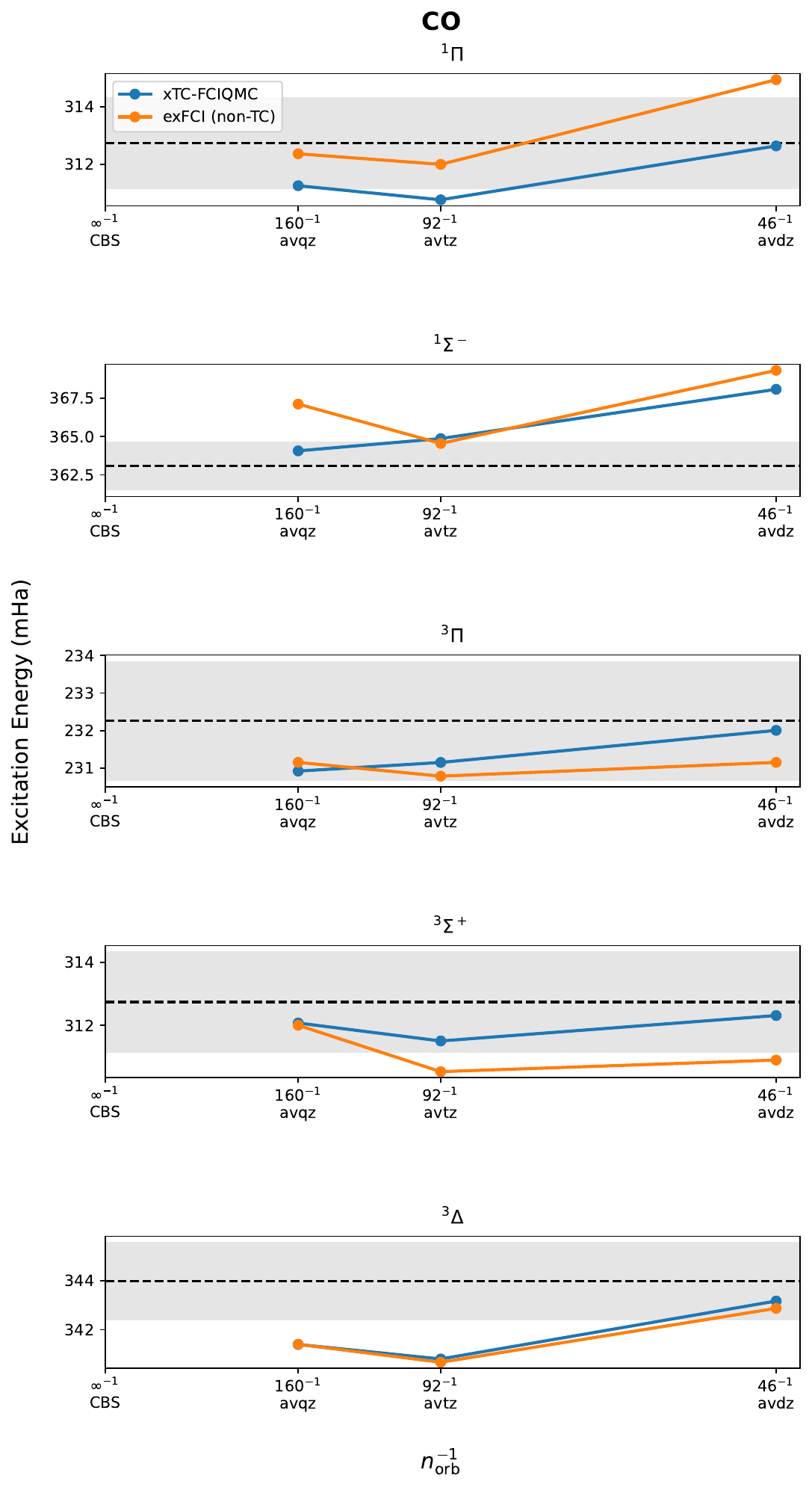}
    \caption{
        Excitation energies for carbon monoxide compared with experiment\supercite{NielsenTransitionMoments,huberConstants1979}.
    }
    \label{fig:excited-states-co}
\end{figure}

\begin{figure}[htbp]
    \centering
    \includegraphics[width=0.5\textwidth]{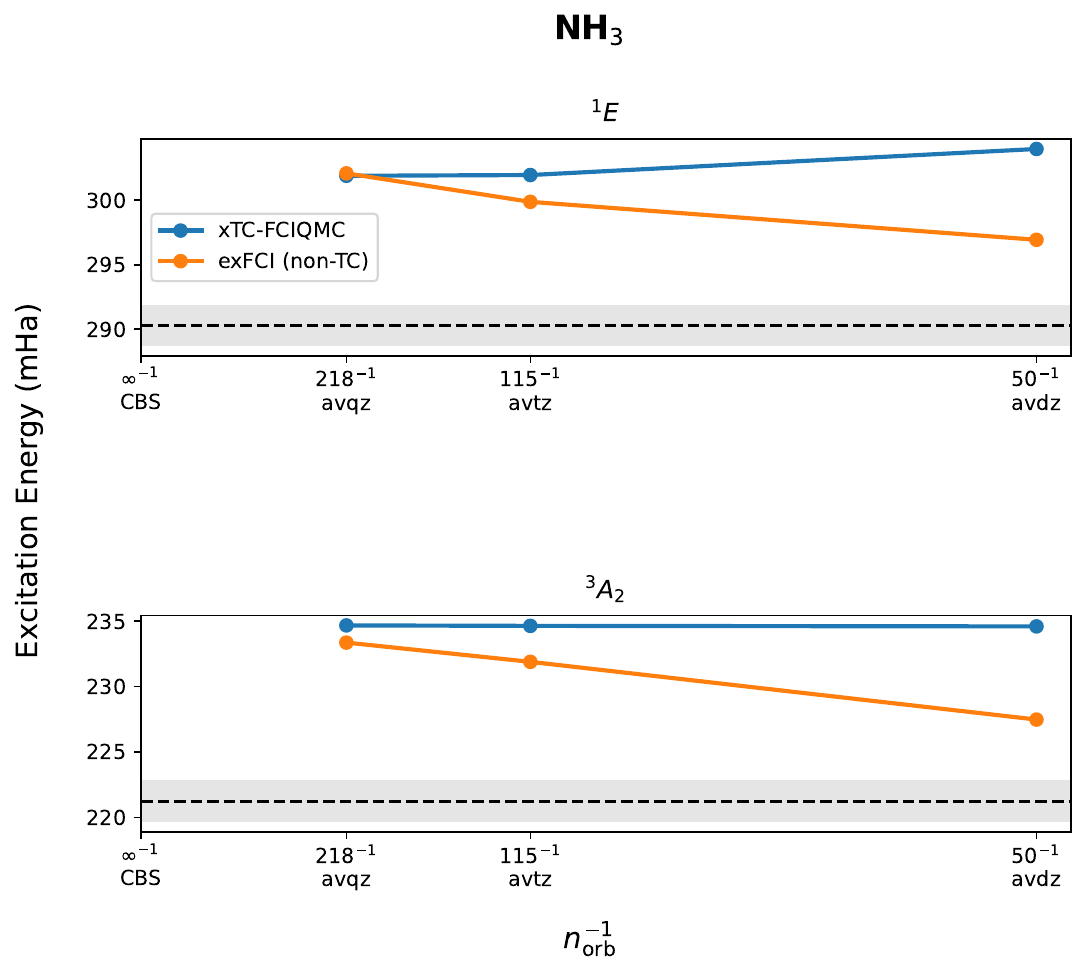}
    \caption{
        Excitation energies for ammonia compared with experiment\supercite{SkerbeleElectronImpact,HarshbargerRelativeIntensities}.
    }
    \label{fig:excited-states-nh3}
\end{figure}

Figures \ref{fig:excited-states-n2}, \ref{fig:excited-states-co} and \ref{fig:excited-states-nh3} show the excitation energies as a function of the reciprocal number of basis functions, $n_\mathrm{orb}^{-1}$ for N$_2$, CO, and NH$_3$ respectively. Experimental values\supercite{oddershedeComparison1985,huberConstants1979,NielsenTransitionMoments,SkerbeleElectronImpact,HarshbargerRelativeIntensities} are included as dashed lines, around which chemical accuracy ($\pm 1.6$ mHa) is shaded grey. Non-TC extrapolated FCI results from reference \cite{loosMountaineering2018} are included for reference. We use the abbreviation av$X$z to refer to basis set \avxz{$X$}.\supercite{avtz}

On the whole, we find that the TC method at \avdz yields results comparable to extrapolated FCI at \avqz, indicating much faster basis-set convergence. The TC excitation energies also converge more monotonically when compared against non-TC-FCI, with the difference between \avtz and \avqz being small.

We also note that the ammonia excitation energies were treated as vertical excitations, which is the likely cause for the large discrepancy compared to experiment. We note that non-TC FCI behaves similarly, but does not converge as rapidly as our TC results.



\section{Conclusion and Outlook}

We have presented a framework for using the transcorrelated method for solving problems of strongly multireference character. We find challenging problems like the nitrogen dissociation curve may be solved by modifying the ansatz for $\Phi$ with a multireference CI expansion, and appropriately changing the corresponding 1RDM for use in the xTC approximation. Particularly effective choices for $\Phi$ are FCIQMC- generate and CASSCF wavefunctions. Using a small FCIQMC calculation has already been found to effectively solve the issues experienced with a HF ansatz, and its use with effective core potentials to efficienctly describe the core region with TC has already been explored.\supercite{simulaEcp} Furthermore, with this methodology we can tailor our TC Hamiltonian to target specific states of interest. Using state-averaged CASSCF, we find that this approach yields accurate results for excitation energies of the set of molecules considered in this study.

One key outlook is to study a wider array of complex problems, including larger molecules, periodic solids, and higher-energy excitations needing orthogonalisation against the ground state. Further studies would also include continuing the procedure self-consistently. That is, after a TC-FCIQMC calculation, we can reuse the CI vector as the $\Phi$ ansatz for a subsequent run until the result is stable. Similarly, this would prove a promising start for a TC-MCSCF method wherein orbitals, CI coefficients and Jastrow factor parameters are all optimised simultaneously. Other interesting directions include state-averaging the Jastrow factors, and improvements to the FCIQMC-Jastrow factor, such as imaginary-time averaging.

\section{Acknowledgements}

The authors gratefully acknowledge the support of the Max Planck Society.

\bibliography{bibliography}

\end{document}